\documentclass[lettersize,journal]{IEEEtran}
\usepackage{amsmath,amsfonts}
\usepackage{array}
\usepackage[caption=false,font=normalsize,labelfont=sf,textfont=sf]{subfig}
\usepackage{textcomp}
\usepackage{stfloats}
\usepackage{url}
\usepackage{verbatim}
\usepackage{graphicx}
\usepackage{cite}
\usepackage{float}
\usepackage{algorithm,algpseudocode}
\algnewcommand{\Inputs}[1]{%
  \State \textbf{Inputs:}
  \Statex \hspace*{\algorithmicindent}\parbox[t]{.8\linewidth}{\raggedright #1}
}
\algnewcommand{\Initialize}[1]{%
  \State \textbf{Initialize:}
  \Statex \hspace*{\algorithmicindent}\parbox[t]{.8\linewidth}{\raggedright #1}
}

\usepackage{hyperref}
\usepackage{multirow}

\begin{document}

\title{PS-Net: Learned Partially Separable Model for Dynamic MR Imaging}

\author{Chentao Cao, Zhuo-Xu Cui, Qingyong Zhu, Congcong Liu, Dong Liang, \IEEEmembership{Senior Member, IEEE}, \\Yanjie Zhu, \IEEEmembership{Member, IEEE}
\thanks{Corresponding author: yj.zhu@siat.ac.cn (Yanjie Zhu)}
\thanks{Chentao Cao and Zhuo-Xu Cui contributed equally to this manuscript}
\thanks{Chentao Cao, Congcong Liu and Yanjie Zhu are with Lauterbur Research Center for Biomedical Imaging, Shenzhen Institutes of Advanced Technology, Chinese Academy of Sciences, Shenzhen, China (e-mail: ct.cao@siat.ac.cn; cc.liu@siat.ac.cn; yj.zhu@siat.ac.cn)}
\thanks{Chentao Cao and Congcong Liu are also with University of Chinese Academy of Sciences, Beijing, China}
\thanks{Yanjie Zhu is also with National Center for Applied Mathematics Shenzhen (NCAMS), Shenzhen, China}
\thanks{Zhuo-Xu Cui, Qingyong Zhu, and Dong Liang are with Research Center for Medical AI, Shenzhen Institutes of Advanced Technology, Chinese Academy of Sciences, Shenzhen, China (e-mail: zx.cui@siat.ac.cn; qy.zhu@siat.ac.cn; dong.liang@siat.ac.cn)}
\thanks{Code available: \href{https://github.com/Aboriginer/PS-Net}{https://github.com/Aboriginer/PS-Net}.}
}

\markboth{Journal of \LaTeX\ Class Files,~Vol.~14, No.~8, August~2021}%
{Shell \MakeLowercase{\textit{et al.}}: A Sample Article Using IEEEtran.cls for IEEE Journals}

\IEEEpubid{0000--0000/00\$00.00~\copyright~2021 IEEE}

\maketitle

\begin{abstract}
Deep learning methods driven by the low-rank regularization have achieved attractive performance in dynamic magnetic resonance (MR) imaging. However, most of these methods represent low-rank prior by hand-crafted nuclear norm, which cannot accurately approximate the low-rank prior over the entire dataset through a fixed regularization parameter. In this paper, we propose a learned low-rank method for dynamic MR imaging. In particular, we unrolled the semi-quadratic splitting method (HQS) algorithm for the partially separable (PS) model to a network, in which the low-rank is adaptively characterized by a learnable null-space transform. Experiments on the cardiac cine dataset show that the proposed model outperforms the state-of-the-art compressed sensing (CS) methods and existing deep learning methods both quantitatively and qualitatively.
\end{abstract}


\section{Introduction}
\IEEEPARstart{D}{ynamic} magnetic resonance imaging (MRI) plays a vital role in cardiac imaging because it can reveal information in spatial anatomy and temporal motion dimensions. However, the inherent long scan time of MRI limits its spatial and temporal resolutions. Therefore, fast dynamic MRI via highly undersampling k-space data generates great research interest. 

With the rise of deep learning, deep learning based MR reconstruction methods have been proposed\cite{MoreTime1, ImageReconstruction} and have shown impressive results.  These methods can be roughly divided into two categories: the end-to-end methods\cite{zhu2018image, Wang, kwon, Han, DC-CNN} and the model-driven methods\cite{ADMMNet,Qin, DLespirit}. The end-to-end strategy uses the neural network to directly learn the mapping from undersampling k-space data to the  label images. It usually requires a large amount of training data, which is difficult to obtain in MR imaging. The model-driven strategy is also called the unrolling-based method. It unrolls the iterative solving step of the compressed sensing (CS) model into a neural network, with learnable regularization and sparsifying transforms, etc. in the model. Compared to end-to-end methods, unrolling-based approaches are more popular in MR reconstruction\cite{prevalent, ramzi2022nc, eo2018kiki} because they require less training data and less GPU resources\cite{cheng2019model}.

Several unrolling-based methods have been proposed for fast dynamic MRI \cite{SLR-Net, modl, L+S-Net, LDC, Manifold, SVTTMI, chen2018novel, Liao, ramzi2022nc, zhang2020deep, bustin2020compressed}. Most of them employ low-rank prior to characterize the correlations among image frames. Low-rank prior is often approximated via the nuclear norm minimization, which is solved by the singular value soft threshold (SVT) with a hand-crafted cutoff parameter\cite{SLR-Net, modl, L+S-Net, SVTTMI, Liao}. When unrolled into a network, it’s straightforward to learn the optimized cutoff parameter of SVT, i.e., the threshold, instead of empirical choosing. However, SVT loses small components and may cause image blurring. In addition, due to its limited representation ability of low-rank and only relaxing low degree of freedom for learning  cutoff parameters, these SVT-learning based methods result in the inadequate characterization of the image low-rank even if the threshold is optimized through learning. In the case of highly accelerated cardiac dynamic imaging, the unrolling-based methods with this type of learnable low-rank are prone to suffer from aliasing artifacts, especially in systole. In CS, another way of unrolling methods is learning the nonlinear transforms via the network to sparsify images more effectively. Since the transforms are adaptively learned, the resulting sparse representation is more accurate than the fixed-base transforms. Inspired by this, a learnable low-rank representation may also enhance the portrayal of the low-rankness in the image than the nuclear norm.
\IEEEpubidadjcol

In this paper, we proposed a partially separable network (PS-Net) to more adaptively characterize the low-rank prior through a learnable null space transform. More specifically, since dynamic MRI can be represented in a partially separable (PS) model \cite{PS-model}, a null space filter can be used to annihilate the PS model according to the classical Prony’s results \cite{Prony's-result}. Then, the low rankness can be implicitly characterized by an adaptively learned null space transform. We combined this low-rank constraint with a sparse constraint under the framework of CS. The corresponding optimization problem was solved in an iterative form with the semi-quadratic splitting method (HQS). The iterative steps were unrolled into a network, dubbed PS-Net. All the regularization parameters and null space transforms are set as learnable in the PS-Net. Our contribution can be summarized as follows:  
\begin{enumerate}
  \item The low rankness of dynamic MRI is first represented by a learnable null space transform, which is more adaptive and accurate than the nuclear norm representation.     
  \item Different from previous learnable transform methods, the null space transform in the PS-Net can be formulated as a Hankel matrix multiplication, which is equivalent to the convolution operation and can be naturally unrolled into a convolutional network module. 
  \item The reconstruction quality of PS-Net is better than the state-of-the-art methods with improved robustness. Moreover, less overlapping of adjacent image frames occurs at high acceleration rates.
\end{enumerate}

The following sections of the paper are organized as follows: Section II introduces the proposed method, Section III describes the experimental results, Section IV provides further discussion, and Section V gives the conclusion.

\section{Methods}
\subsection{Problem Formulation}
In order to speed up the acquisition of dynamic MRI, the k-space data are usually undersampled. Thereby, the reconstruction method aims to recover high-quality dynamic MR images from the undersampled k-space data. Under the CS framework, dynamic MRI reconstruction can be described as the following inverse problem:
\begin{equation}
\gamma^{*}=\underset{\gamma}{\arg \min } \frac{1}{2}\sum_{i=1}^m\|M Fc_i \gamma-y\|_{F}^{2}+ R(\gamma),
\end{equation}
where $\gamma$ is the dynamic MR image to be reconstructed, $y$ represents the undersampling k-space data, $M$ is the undersampling  mask, $F$ denotes Fourier transform, $c_i$ is the diagonal matrix transformed from the sensitivity of the $i$th coil, $m$ indicates the coil number, and $R(\gamma)$ is the regularization term. $\gamma, y \in \mathbb{C}^{N_x\times N_y \times N_t}$. $N_x$ and $N_y$ denote the length and width of the dynamic MR images, and $N_t$ is the number of frames in the time direction.

\subsection{Partially Separable Model}
In dynamic MRI, the low-rank and sparse properties of the images are often explored as the regularization terms\cite{kt-SLR, Low-Rank, majumdar2015improving}:
\begin{equation}
\begin{gathered}
\gamma^{*}=\underset{\gamma}{\arg \min } \frac{1}{2}\sum_{i=1}^m\|M Fc_i \gamma-y\|_{F}^{2}\\
s.t.~\left\{
\begin{aligned}&\|R_1(\gamma)\|_0 \leq S\\
~&\|R_2(\gamma)\|_* \leq L\end{aligned}
\right.,
\label{hand-craft}
\end{gathered}
\end{equation}
where $R_1(\gamma)$ is the sparse constraint and $R_2(\gamma)$ is the low-rank constraint.

In the previous study, Liang proposed the partially separable model\cite{PS-model} to characterize low-rank, which represents the dynamic image as the sum of the products of temporal basis functions and spatial basis functions:
\begin{equation}
\gamma(r, t)=\sum_{l=1}^{L} c_{l}(r) \phi_{l}(t),
\end{equation}
where $\phi_l(t)$ is the temporal basis functions, $c_l(r)$ is the spatial basis functions, and $\gamma(r, t)$ is the image to be reconstructed. $r$ is the coordinate in image domain and $t$ is the coordinate in time direction.

\subsection{The Proposed Method: Deep Partially Separable Modelling for Dynamic MRI}
\subsubsection{The PS annihilation module}
Based on PS model, we proposed an adaptive module to characterize the low rankness of dynamic MRI. According to the classical Prony’s results\cite{Prony's-result}, there exists a $L$-tap filter $h_{ps}(r, t)$ annihilates the PS model $\gamma(r, t)$:
\begin{equation}
(\gamma * h_{ps})(r, t)=0, \quad \forall r, t,
\label{PS annihilation}
\end{equation}
where $h_{ps}(r, t)$ has the following form:
\begin{equation}
h_{ps}(r, t)=\prod_{l=1}^{L}\left(1-\phi_{l}(t)^{\frac{1}{t}} r^{-1}\right).
\end{equation}

By the equivalence relation of Hankel matrix multiplication and convolution, \eqref{PS annihilation} is equivalent to: 
\begin{equation}
H(\gamma(r, t))h_{ps}(r, t)=0, \quad \forall r, t,
\end{equation}
where $H(\gamma)$ is a linear convolution matrix with Hankel-structure in the $N_t$ direction and $h_{ps}(r, t)$ is also called the null space filter. Then the low rankness can be implicitly characterized via this null space filter.

\subsubsection{The sparse module}
Since the signal intensities of adjacent pixels in dynamic MR images are close to each other, the sparse property is satisfied after performing the gradient on the image\cite{Deep-SLR}:
\begin{equation}
(\nabla \gamma(r)) \mu(r) = 0, \quad \forall r,
\label{sparse}
\end{equation}
where $\mu$ is a set of bandlimited functions, $\mu \in \mathbb{C}^{N_x\times N_y}$. By the convolution theorem, \eqref{sparse} can be equivalently expressed as:
\begin{equation}
(\widehat{\nabla \gamma} * h_s)(w) = 0, \quad \forall w,
\label{conv}
\end{equation}
where $\widehat{\nabla \gamma}$ represents the Fourier coefficients of $\nabla \gamma(r)$ (along the $N_xN_y$ direction), $w$ is the coordinate in frequency domain, $h_s(w)$ is the frequency domain signal obtained by the Fourier transform of $\mu(r)$. By the equivalence of convolution and Hankel matrix multiplication, \eqref{conv} can be rewritten as:
\begin{equation}
H(\widehat{\nabla \gamma}(w))h_s(w)=0, \quad \forall w,
\end{equation}
where $H(\widehat{\nabla\gamma}(w))$ is the Hankel-structured matrix in spatial dimension.

\subsubsection{The Proposed Model}
With the above PS annihilation module and sparse module, we formulate the dynamic MRI reconstruction problem as the following unconstrained two-norm minimization problem:
\begin{equation}
\begin{gathered}
\underset{\gamma}{\arg \min } \frac{1}{2}\sum_{i=1}^m\|M Fc_i \gamma-y\|_{F}^{2} + \lambda_1\sum_{i=1}^T\|H_1(\widehat{\nabla\gamma_i})h_{si}\|^2_F \\
+ \lambda_2||H_2(\gamma)h_{ps}||^2_F,
\label{unconstrained}
\end{gathered}
\end{equation}
where $||H_1(\widehat{\nabla\gamma_i})h_{si}||^2_F$ represents the sparse regularizer, $T$ is the number of time frames, and $||H_2(\gamma)h_{ps}||^2_F$ is the null space annihilation  regularizer, which is derived by the partially separable property. $\widehat{\nabla\gamma}=[\widehat{\nabla\gamma_1}, \widehat{\nabla\gamma_2}, ..., \widehat{\nabla\gamma_T}], h_{s}=[h_{s1},h_{s2}, ..., h_{sT}]$, $\lambda_1$ and $\lambda_2$ are the corresponding regularization parameters. To solve the optimization problem in \eqref{unconstrained}, we introduce auxiliary variables $x_i$, $\widehat{U_i}$, $Z$. The reformed optimization problem is as follows:
\begin{equation}
\begin{gathered}
\min _{\gamma, x_i \widehat{U_i}, Z} \frac{1}{2}\sum_{i=1}^m\|M Fx_i-y\|_{F}^{2}+\lambda_{1}\sum_{i=1}^T\|H_1(\widehat{U_i}) h_{si}\|_{F}^{2}\\
+\lambda_{2}\|H_2(Z) h_{ps}\|_{F}^{2} \\
s.t.~~~~x_i = c_i\gamma,~\widehat{U_i} = \widehat{\nabla\gamma_i},~Z = \gamma,
\label{reformed optimization}
\end{gathered}
\end{equation}
By substituting $
\widehat{\nabla\gamma_i} = \left(\begin{array}{l}
i w_{x} F\left(\gamma_i\right) \\
i w_{y} F\left(\gamma_i\right)
\end{array}\right)$, where $w_x$ and $w_y$ denote the k-space coordinates in $x$ and $y$ direction, respectively. The penalty function of \eqref{reformed optimization} can be written as:
\begin{equation}
\begin{gathered}
\min _{\gamma, x_i \widehat{U_i}, Z} \frac{1}{2}\sum_{i=1}^m\|M Fx_i-y\|_{F}^{2}+\lambda_{1}\sum_{i=1}^T\|H_1(\widehat{U_i}) h_{si}\|_{F}^{2}\\
+\lambda_{2}\|H_2(Z) h_{ps}\|_{F}^{2} + \frac{\rho_0}{2}\sum_{i=1}^m\|x_i-c_i\gamma\|_{F}^{2}\\
+ \frac{\rho_{1}}{2}\sum_{i=1}^T\|\widehat{U_i}-\left(\begin{array}{l}
i w_{x} F(\gamma_i) \\
i w_{y} F(\gamma_i)
\end{array}\right)\|_F^{2} + \frac{\rho_{2}}{2}\|Z-\gamma\|_{F}^{2}.
\end{gathered}
\end{equation}
Based on the semi-quadratic splitting method (HQS), we can construct an iterative solution algorithm for the above penalty function:
\begin{equation}
\left\{\begin{aligned}
\widehat{U_i}^{k}=& \min _{\widehat{U_i}}\frac{\rho_{1}}{2}\|\widehat{U_i}-\left(\begin{array}{l}
i w_{x} F(\gamma_i^{k-1}) \\
i w_{y} F(\gamma_i^{k-1})
\end{array}\right)\|_{F}^{2} \\
&+ \lambda_{1}\|H_1(\widehat{U_i}) h_{si}\|_{F}^{2}\\
Z^{k}=& \min _{\mathrm{Z}} \frac{\rho_{2}}{2}\|Z-\gamma^{k-1}\|_{F}^{2}+\lambda_{2}\left\|H_2(Z) h_{ps}\right\|_{F}^{2}\\
x_i^k=&\min _{x_i} \frac{1}{2}\|MFx_i-y\|_{F}^{2}+\frac{\rho_0}{2}\|x_i-c_i\gamma^{k-1}\|_{F}^{2}\\
\gamma^{k}=& \min _{\gamma}
\frac{\rho_{1}}{2}\sum_{i=1}^T\|\widehat{U_i}^{k}-\left(\begin{array}{c}
i w_{x} F(\gamma_i) \\
i w_{y} F(\gamma_i)
\end{array}\right)\|_{F}^{2}\\
&+\frac{\rho_0}{2}\sum_{i=1}^m\|x_i^{k}-c_i\gamma\|_{F}^{2}+\frac{\rho_{2}}{2}\|Z^{k}-\gamma\|_{F}^{2}
\end{aligned}\right..
\label{subproblems}
\end{equation}
By the first-order optimization condition, it reads:
\begin{equation}
\left\{\begin{aligned}
\widehat{U_i}^{k}=&\left(I-\frac{2 \lambda_{1}}{\rho_{1}} H_{h_{si}}^{T} H_{h_{si}}\right)\left(\begin{array}{c}
i w_{x} F(\gamma^{k-1}_i) \\
i w_{y} F(\gamma^{k-1}_i)
\end{array}\right) \\
Z^{k}=&\left(I-\frac{2 \lambda_{2}}{\rho_{2}} H_{h_{ps}}^{T} H_{h_{ps}}\right) \gamma^{k-1} \\
x_i^k=& F^{-1}\left(\frac{My+\rho_0F(c_i\gamma^{k-1})}{M+\rho_0}\right)\\
\gamma^{k}=& F^{-1}\left(\frac{\rho_{1}\left(\left(i w_{x}\right)^{*},\left(i w_{y}\right)^{*}\right) \widehat{U}^{k}}{M+\rho_0+\rho_{1}(w_xw_{x}^*+w_{y}w_{y}^*)+\rho_{2}}\right)+\\
&F^{-1}\left(\frac{M y+\rho_0F(\sum_{i=1}^mc_i^*x_i^{k}) +\rho_{2} F(Z^{k})}{{M+\rho_0+\rho_{1}(w_xw_{x}^*+w_{y}w_{y}^*)+\rho_{2}}}\right)
\end{aligned}\right..
\label{iteratively results}
\end{equation}

\begin{figure*}
\centerline{\includegraphics[width=18cm]{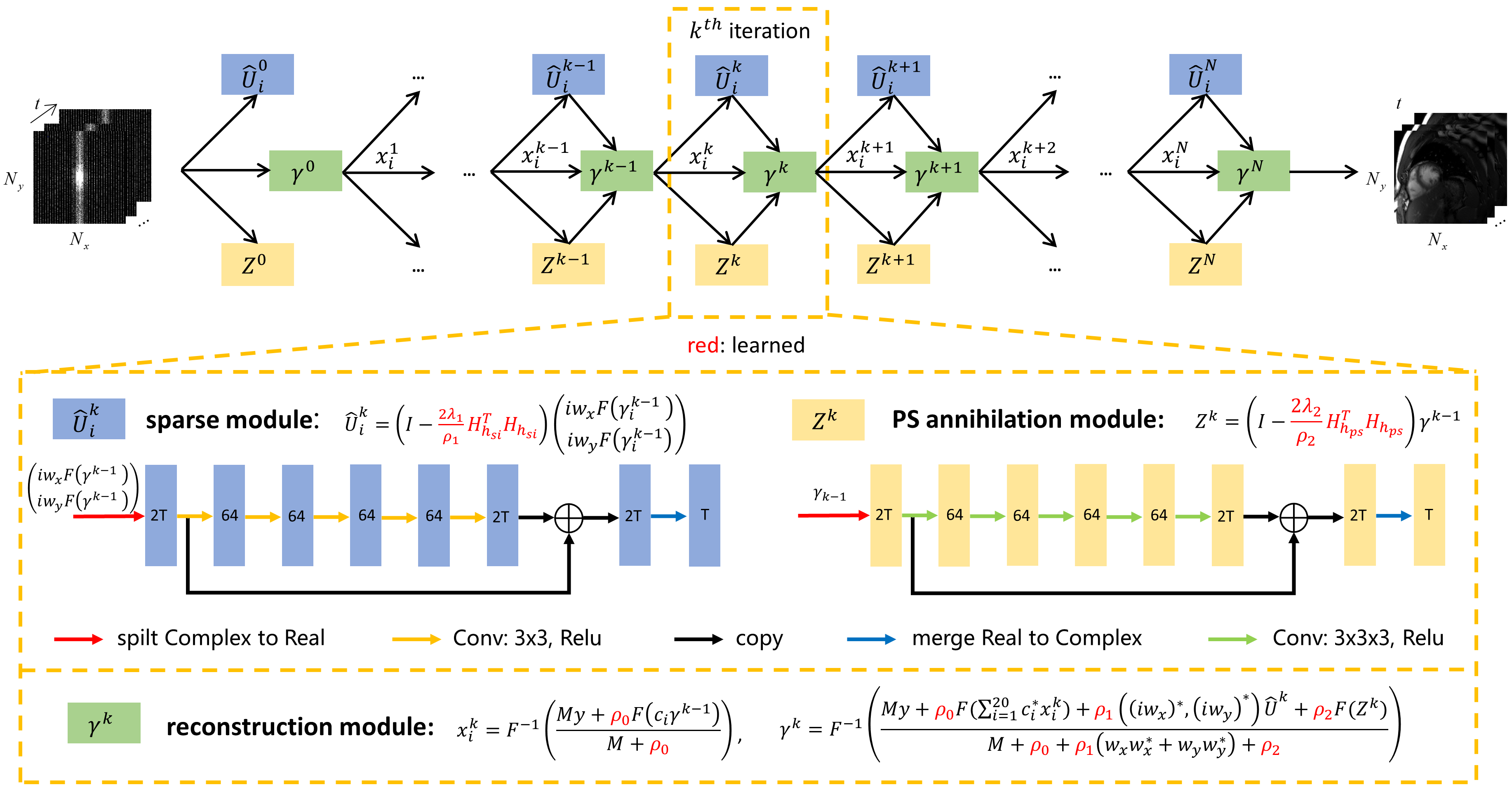}}
\caption{The network architecture of PS-Net. The input of the network is k-space data and the output is dynamic MR images. PS-Net has three modules: the sparse module, the PS annihilation module and the reconstruction module. The three modules are iterated 10 times ($N=9$). The sparse and PS annihilation modules are implemented by convolutional residual network with $64$ channels in the first four layers and $T$ channels in the fifth layer. $T$ indicates the number of frames of the input data in the time direction ($T=18$). The sparse module convolves along the $N_xN_y$ direction and the PS annihilation module convolves along the $t$ direction. The data are split into real input convolutional layers before each iteration and reverted to complex numbers at the end.}
\label{network structure}
\end{figure*}

$\gamma$ can be obtained by iteratively solving the four sub-problems of \eqref{iteratively results}, but iteratively solving \eqref{iteratively results} directly requires adjusting many parameters ($\lambda_1, \lambda_2, \rho_1, \rho_2$).

\subsubsection{The Proposed Network PS-Net}
We solve the above problem by unrolling \eqref{iteratively results} into PS-Net. The above four sub-problems correspond to three modules in the network: sparse module $S$ (corresponding to the $\widehat{U}_i^k$ sub-problem), PS Annihilation module $PS$ (corresponding to the $Z^k$ sub-problem), and reconstruction module $R$ (corresponding to the $x_i$ and $\gamma^k$ sub-problems). The overall structure of PS-Net is shown in Fig. \ref{network structure}. There are 10 iterations in the network. The regularization parameters are learned by the network adaptively. The detailed structures of the three models in the network is shown below.
\begin{itemize}
  \item \textbf{The sparse module $S$}: Since the Hankel matrix multiplication and the convolution are equivalent, the $\widehat{U}_i^k$ sub-problem is solved via the convolution layer in the network. The convolution layer is set to 2D convolution along the $N_x$ and $N_y$ directions because the gradient image of dynamic MRI satisfies the sparse property along these two directions. The number of convolution layers is 5 and the number of convolution channels is 64. We split the real and imaginary parts of the input data into two channels, which means that the dimension size in the time direction changes from $T$ to $2T$.
  \item \textbf{The PS annihilation module $PS$}: The PS annihilation layer characterizes the low-rank prior by a learnable null space filter mapping. Similarly, $Z^k$ is implemented equivalently through the convolution layers. Considering that the image domain has redundancy \cite{Deep-SLR}, we consider not only the convolution in the time direction but also along the $N_x$ and $N_y$ directions. The numbers of convolution layers and convolution channels are the same as the sparse module. The input data is also divided into real and imaginary channels.
  \item \textbf{The reconstruction module $R$}: According to the first-order optimization condition, the minimal value of the energy function with respect to $\gamma$ can be solved by \eqref{iteratively results}, thus ensuring data consistency.
\end{itemize}

The pseudo-code for the network unrolling solution is shown in Alg. 1.

\subsubsection{Network configuration}
PS-Net reconstructs the image via supervised learning. The label of PS-Net is the fully sampled dynamic MR image. Input data is generated by artificially adding the undersampling mask to the label. The network's output is the reconstructed dynamic MR image, using the minimum mean square error as the loss function, which can be expressed as:
\begin{equation}
Loss(\gamma^{*})=\frac{1}{N_{samples}} \sum_{i=1}^{N_{samples}}\left(\gamma^{*}-\gamma_{ref}\right)^{2}_F,
\end{equation}
where $\gamma^*$ is the output, $\gamma_{ref}$ is the label, $N_{samples}$ is the total number of samples. The activation function in the network is handled by Relu\cite{ReLU}, and Adam has been selected as the optimizer\cite{kingma2014adam}. All hyper-parameters in the network are initialized to 1 ($\lambda_1, \lambda_2, \rho_1, \rho_2$). We set the epoch of network training to 50. The initial value of the learning rate is 0.001; the learning rate delay is 0.95; the experimental environment is tensorflow2.7\cite{Tensorflow}, cuda11.3, ubuntu20.04, and the graphics card is NVIDIA A100 Tensor Core GPU.

\begin{algorithm}
  \caption{PS-Net}
  \begin{algorithmic}[1]
    \Inputs{$M$: The undersampling mask \\ $F$: The 2D spatial Fourier operator \\ $y$: The undersampling k-space data}
    \Initialize{\strut$\widehat \gamma^0=F^{-1}(y), U^0_i = \widehat{\nabla\gamma_i},Z^0=F^{-1}(y)$\\
    $\lambda_1,\lambda_2,\rho_1,\rho_2=1$}
    \For{$k = 1$ to $N_{niter-1}$}
      \State $\widehat{U}_i^{k}=\left(I-\frac{2 \lambda_{1}}{\rho_{1}} H_{h_{si}}^{T} H_{h_{si}}\right)\left(\begin{array}{l}
i w_{x} F\left(\gamma_i^{k-1}\right) \\
i w_{y} F\left(\gamma_i^{k-1}\right)
\end{array}\right) $
      \State $Z^{k}=\left(I-\frac{2 \lambda_{2}}{\rho_{2}} H_{h_{ps}}^{T} H_{h_{ps}}\right) \gamma^{k-1}$
      \State $x_i^k=F^{-1}(\frac{My+\rho_0F(c_i\gamma^{k-1}}{M+\rho_0})$
      \State $\gamma^{k}=F^{-1}\left(\frac{\rho_{1}\left(\left(i w_{x}\right)^{*},\left(i w_{y}\right)^{*}\right)\widehat{U}^{k}}{M+\rho_0+\rho_{1}(w_xw_{x}^{*}+w_yw_{y}^{*})+\rho_{2}}\right)$
      \State $+F^{-1}\left(\frac{M y+\rho_0F(\sum_{i=1}^nc_i^*x_i^{k})+\rho_{2} F(Z^{k})}{M+\rho_0+\rho_{1}(w_xw_{x}^{*}+w_yw_{y}^{*})+\rho_{2}}\right)
      $
    \EndFor\\
    \Return $\gamma^{N_{iter-1}}$
  \end{algorithmic}
\end{algorithm}

\section{Experimental results}
\subsection{Setup}
\subsubsection{The cardiac dataset}
The cardiac cine dataset was acquired from 30 individuals using a 3T MRI scanner (Trio, Siemens, 20-channel receiver coil). The Ethics Review Committee of the Shenzhen Institute of Advanced Technology, Chinese Academy of Sciences, authorized all scanning studies with the subjects’ informed consent. The imaging parameters were: acquisition matrix = 256$\times$256, slice thickness = 6mm, TR/TE = 3.0ms/1.5ms, and FOV = 330$\times$330mm. Total 356 slices were acquired. PS-Net was performed in both single-channel and multi-channel data, where the single-channel data was obtained by fusing the sensitivity of 20 coils and the coil sensitivity in \eqref{unconstrained} is set to the identity matrix for the single-channel scenario. Data from 25 individuals were randomly selected as the training dataset, and the remaining 5 individuals were used as the test dataset. We cropped the dynamic images along the $x$, $y$, and $t$ direction and used rigid transform cropping for data enhancement. Eventually, the training dataset had 800 images, and the test dataset had 118 images, with a size of 192$\times$192$\times$18.

The data was undersampled by applying undersampling masks on the fully sampled k-space data. Random cartesian undersampling masks\cite{Cartesian-mask, vista} were employed for the fully sampled data. The phase coding direction is undersampled and the frequency coding direction is fully sampled. The input and fully sampled images are coupled one by one as the training pairs.

\subsubsection{Performance evaluation}
The quantitative evaluation metrics include minimum mean square error (MSE), peak signal-to-noise ratio (PSNR), and structural similarity (SSIM):
\begin{equation}
\mathrm{MSE}=\frac{1}{n} \sum_{i=1}^{n}\left(\gamma^{*}-\gamma_{ref}\right)^{2}_2,
\end{equation}
\begin{equation}
\mathrm{PSNR}=20 \log _{10}\left(\frac{max_{\gamma^*}}{\sqrt{MSE}}\right),
\end{equation}
\begin{equation}
\mathrm{SSIM}=l(\gamma^*, \gamma_{ref})\cdot c(\gamma^*, \gamma_{ref}) \cdot s(\gamma^*, \gamma_{ref}),
\end{equation}
where $\gamma^*$ is the output and $\gamma_{ref}$ is the label, $l,c,s$ are three comparison measurements between the samples of $\gamma^*$ and $\gamma$\cite{SSIM}. The smaller MSE, the larger PSNR and SSIM, indicate a better reconstruction.

\subsection{Results}
\subsubsection{Single-channel reconstruction results}
In the single-channel scenario, we compared PS-Net with the traditional compressed sensing method L+S\cite{L+S}, the deep learning end-to-end method CRNN\cite{CRNN}, and the model-driven based method SLR-Net\cite{SLR-Net}.
\begin{figure}
\centerline{\includegraphics[width=9cm]{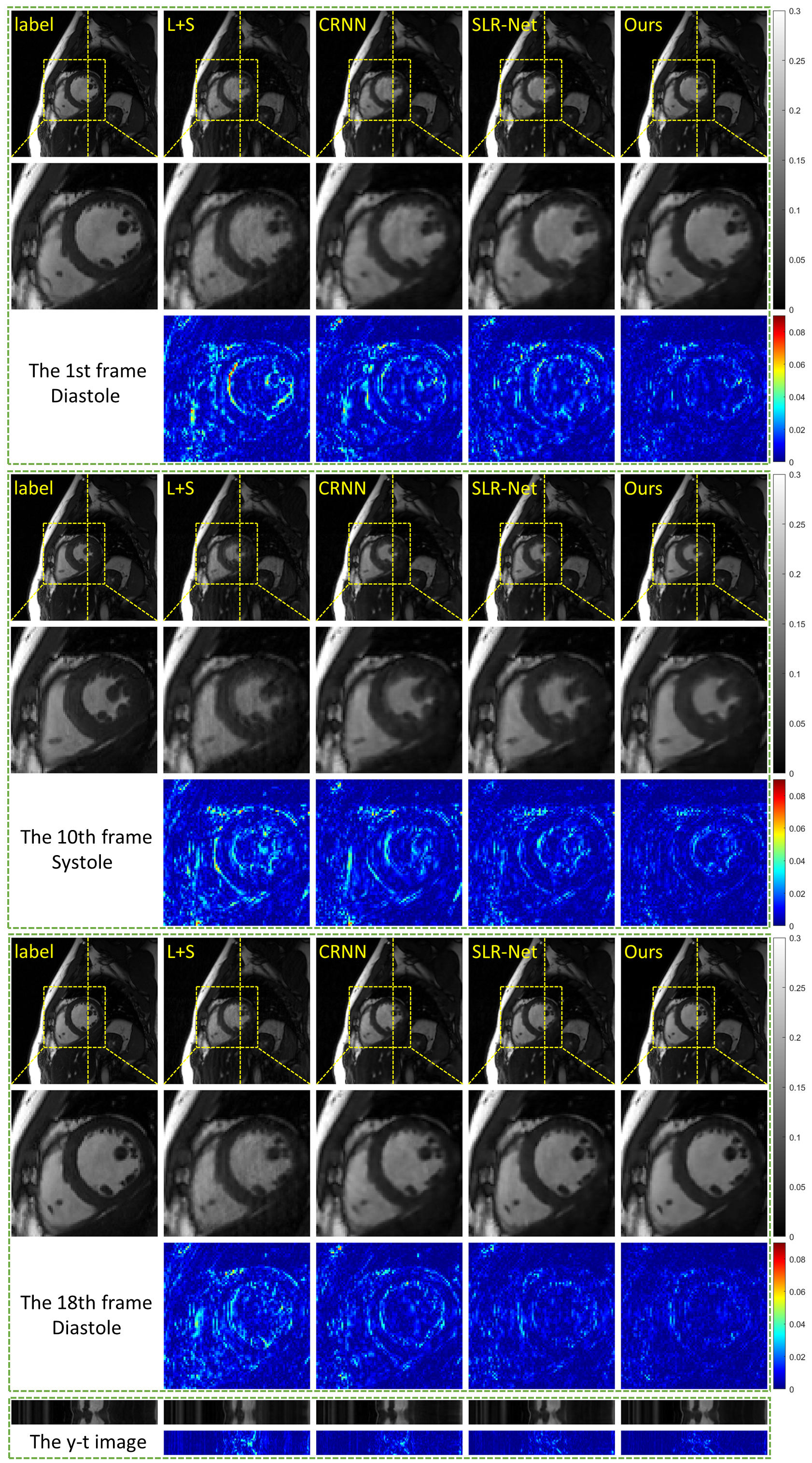}}
\caption{The reconstruction results in the single-channel, 8-fold undersampling scene. From top to bottom (except the y-t image), Fig. \ref{single results} shows the reconstruction results of the 1st frame, the 10th frame, and the 18th frame to represent the dynamic images both in diastole and systole. The first row of each frame shows the reconstruction results of L+S, CRNN, SLR-Net, and PS-Net. The second row shows the zoomed-in view of the ROI region (indicated by the yellow box in the first row), and the third row shows the error map of the ROI region. The y-t image in the bottom shows the 92nd slice along the $y$ and temporal direction. The error range of the error map is [0--0.09].}
\label{single results}
\end{figure}

The reconstruction results of the single-channel data undersampled by 8-fold acceleration are shown in Fig. \ref{single results}. From top to bottom (except the y-t image), Fig. \ref{single results} shows the reconstruction results of the 1st frame, the 10th frame, and the 18th frame to represent the dynamic images both in diastole and systole. The first row of each frame shows the reconstruction results of L+S, CRNN, SLR-Net, and PS-Net. The second row shows the zoomed-in view of the ROI region (indicated by the yellow box in the first row), and the third row shows the error map of the ROI region. The y-t image in the bottom shows the 92nd slice along the $y$ and temporal direction. The error range of the error map is [0--0.09].


\begin{figure*}
\centerline{\includegraphics[width=18.5cm]{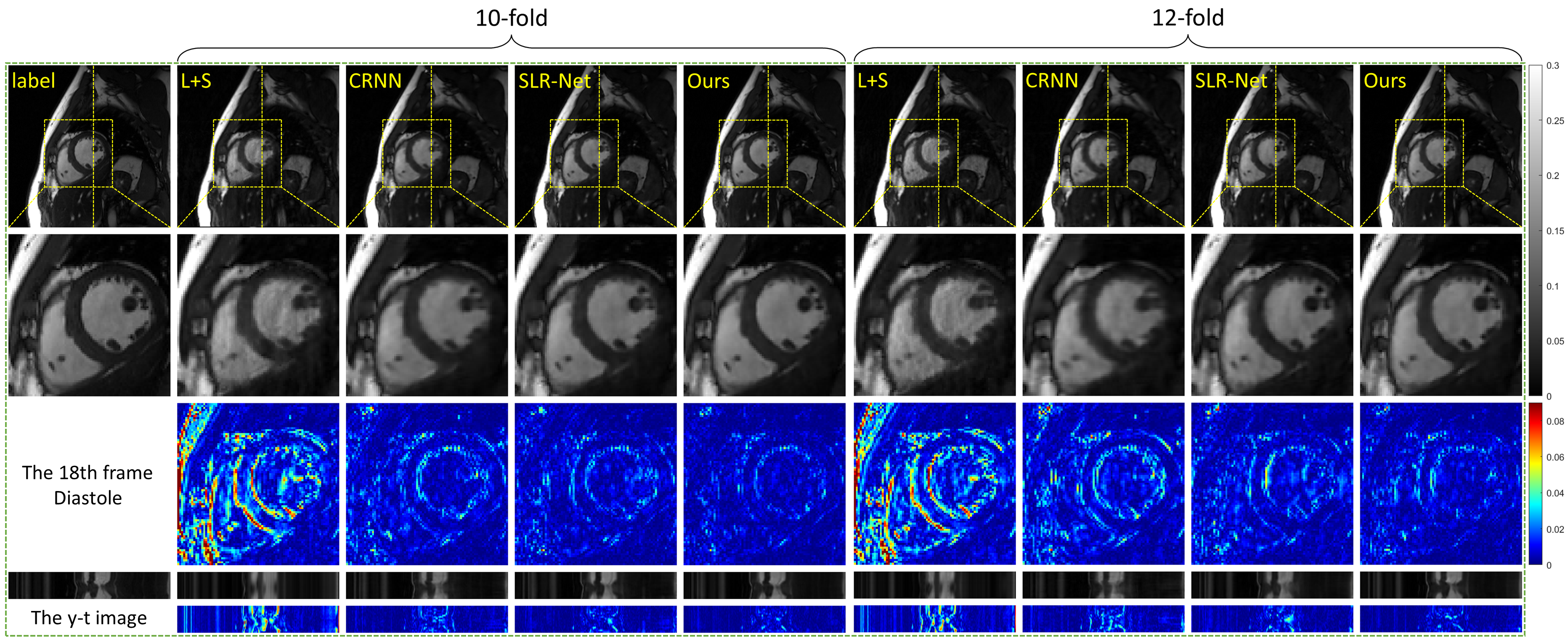}}
\caption{The reconstruction results in the single-channel, 10-fold and 12-fold undersampling scene. The label appears in the first column, followed by 10-fold undersampling reconstruction results in the second to fifth columns, then 12-fold undersampling reconstruction results in the last four columns. The label and the reconstruction results for each approach are displayed in the first row. The second row shows the zoomed-in view of the ROI region (indicated by the yellow box in the first row), and the third row shows the error map of the ROI region. The y-t image in the fourth row shows the 92nd slice along the $y$ and temporal direction. The fifth row shows the error map of the y-t image with an error range of [0--0.09].}
\label{acc 10 12 single results}
\end{figure*}


The enlarged ROI region in the second row of the 1st and the 18th frame (the diastolic phase) suggest that PS-Net clearly reconstructs the tiny structures of papillary muscle without artifacts and outperforms other methods in detail retention and artifact removal. The error map of the reconstruction in the fifth row along the temporal dimension shows that PS-Net also outperforms the other methods in the dynamic map reconstruction. The reconstructed images of the 10th frame (the systolic phase) are not as good as the 1st and the 18th frame. Although all images are blurred due to high undersampling rate in the 10th frame, the PS-Net achieves the lowest errors with no significant deformation appears in the reconstructed results.

Compared with SLR-Net, which learns low-rank by the singular value soft threshold (SVT), we found that PS-Net is more accurate for low-rank learning, indicating that our adaptive low-rank learning strategy outperforms the representing low-rank prior by the hand-crafted nuclear norm.

We also performed the experiments at higher acceleration rates with 10-fold and 12-fold. The results are shown in Fig. \ref{acc 10 12 single results}. PS-Net clearly reconstructs the tiny structures of papillary muscle even at 12-fold acceleration. The quantitative evaluation metrics are shown in Table \ref{tab1 single coil}, and it can be seen that the performance of PS-Net outperforms the state-of-the-art methods.

\begin{table}
\caption{The average values for the quantitative evaluation metrics (MSE, PSNR, and SSIM) of the single-channel testing dataset after it was undersampled by 8-fold, 10-fold and 12-fold.}
\begin{center}
\begin{tabular}{c|cccc}
\hline \hline AF & Methods & MSE(*e-5) & PSNR (dB) & SSIM(*e-2) \\
\hline \multirow{4}{*}{8$\times$}& L+S & 11.54 $\pm$ 5.12 & 39.75 $\pm$ 1.75 & 94.46 $\pm$ 1.68 \\
& CRNN & 5.60 $\pm$ 1.67 & 42.70 $\pm$ 1.24 & 97.07 $\pm$ 0.61 \\
& SLR-Net & 4.82 $\pm$ 1.90 & 43.48 $\pm$ 1.62 & 97.40 $\pm$ 0.73 \\
& PS-Net & \textbf{2.91 $\pm$ 1.25} & \textbf{45.74 $\pm$ 1.78} & \textbf{98.19 $\pm$ 0.64} \\
\hline \multirow{4}{*}{10$\times$}& L+S & 113.26 $\pm$ 25.71 & 29.57 $\pm$ 0.97 & 86.47 $\pm$ 1.41 \\
& CRNN & 7.28 $\pm$ 2.25 & 41.58 $\pm$ 1.31 & 96.49 $\pm$ 0.75 \\
& SLR-Net & 8.07 $\pm$ 2.86 & 41.19 $\pm$ 1.49 & 95.94 $\pm$ 0.94 \\
& PS-Net & \textbf{4.29 $\pm$ 1.70} & \textbf{43.99 $\pm$ 1.65} & \textbf{97.63 $\pm$ 0.77} \\
\hline \multirow{4}{*}{12$\times$}& L+S & 113.42 $\pm$ 25.90 & 29.56 $\pm$ 0.90 & 86.46 $\pm$ 1.45 \\
& CRNN & 11.87 $\pm$ 3.35 & 39.43 $\pm$ 1.21 & 94.57 $\pm$ 0.89 \\
& SLR-Net & 9.42 $\pm$ 3.60 & 40.53 $\pm$ 1.50 & 95.35 $\pm$ 1.05 \\
& PS-Net & \textbf{5.29 $\pm$ 2.09} & \textbf{43.09 $\pm$ 1.67} & \textbf{97.16 $\pm$ 0.89} \\
\hline \hline
\end{tabular}
\label{tab1 single coil}
\end{center}
\end{table}

\subsubsection{Multi-channel reconstruction results}
\begin{figure}
\centerline{\includegraphics[width=9cm]{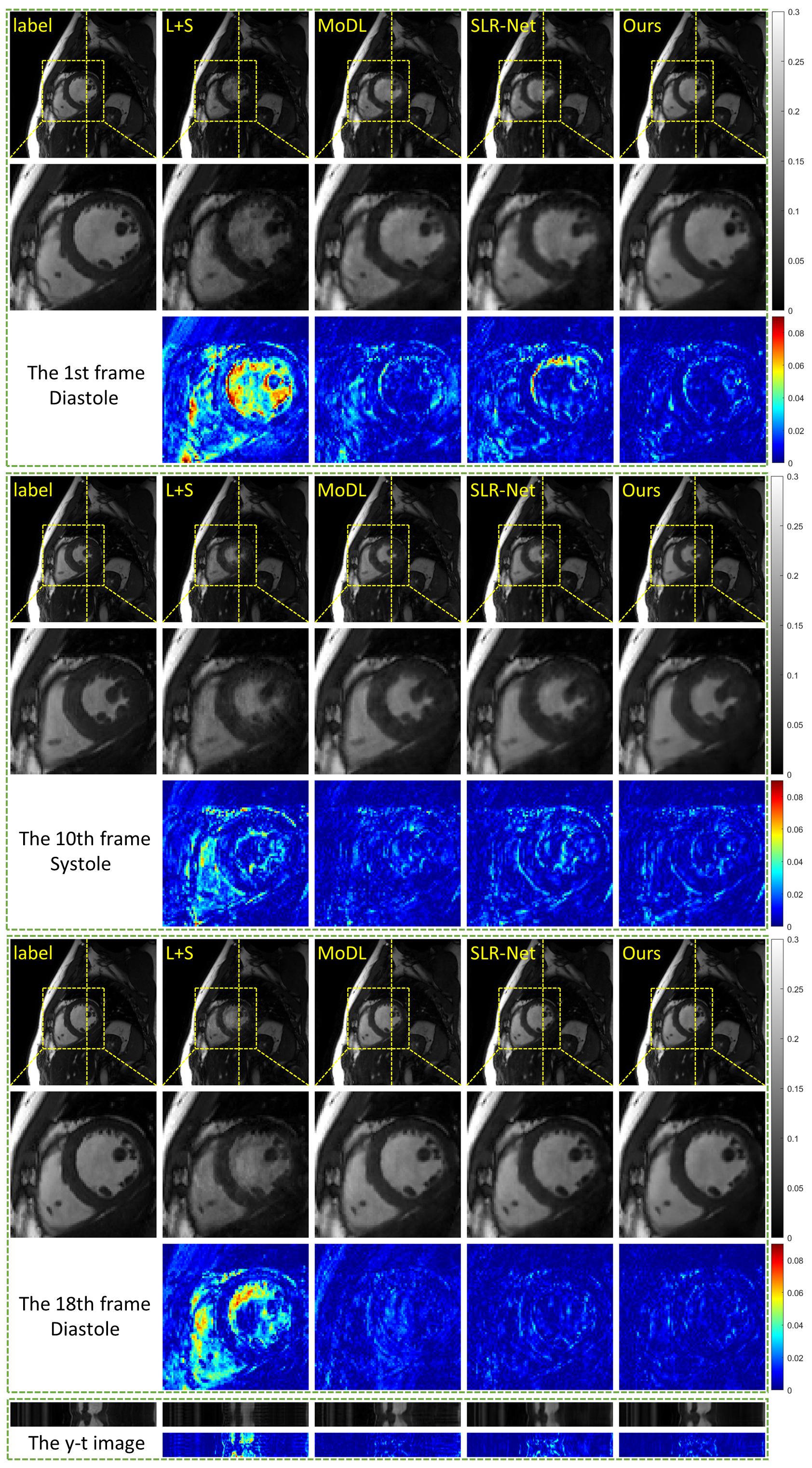}}
\caption{The reconstruction results in the multi-channel, 12-fold undersampling scene. From top to bottom (except the y-t image), Fig. \ref{single results} shows the reconstruction results of the 1st frame, the 10th frame, and the 18th frame to represent the dynamic images both in diastole and systole. The first row of each frame shows the reconstruction results of L+S, MoDL, SLR-Net, and PS-Net. The second row shows the zoomed-in view of the ROI region (indicated by the yellow box in the first row), and the third row shows the error map of the ROI region. The y-t image in the bottom shows the 92nd slice along the $y$ and temporal direction. The error range of the error map is [0--0.09].}
\label{multi-coil}
\end{figure}

In the multi-channel scenario, comparisons were made with the L+S method\cite{L+S}, and the model-driven based methods MoDL\cite{modl} and SLR-Net\cite{SLR-Net}. Because multi-channel data contains redundant information such as coil sensitivity, multi-channel data can be accelerated to a higher rate. The reconstruction results of multi-channel data undersampling with 12-fold acceleration are shown in Fig. \ref{multi-coil}, and the quantitative evaluation metrics are shown in Table \ref{mul acc12}. In the multi-channel case, the reconstruction result of PS-Net outperforms the L+S, MoDL, and SLR-Net in terms of detail description, artifact removal, etc.

\begin{table}
\caption{The average quantitative evaluation metrics (MSE, PSNR, SSIM) of L+S, MoDL, SLR-Net and PS-Net in the multi-channel testing dataset when undersampled by 12-fold.}
\begin{center}
\begin{tabular}{c|cccc}
\hline \hline AF & Methods & MSE(*e-5) & PSNR (dB) & SSIM(*e-2) \\
\hline \multirow{4}{*}{12$\times$}& L+S & 20.89 $\pm$ 13.54 & 37.52 $\pm$ 2.35 & 93.31 $\pm$ 3.47 \\
&MoDL & 6.14 $\pm$ 2.64 & 42.46 $\pm$ 1.67 & 96.29 $\pm$ 0.98 \\
&SLR-Net & 6.75 $\pm$ 3.31 & 42.15 $\pm$ 1.90 & 96.58 $\pm$ 1.08 \\
&PS-Net & \textbf{3.67 $\pm$ 1.89} & \textbf{44.87 $\pm$ 2.07} & \textbf{98.04 $\pm$ 0.85} \\
\hline \hline
\end{tabular}
\label{mul acc12}
\end{center}
\end{table}

\subsubsection{Ablation experiments}

\begin{figure}
\centerline{\includegraphics[width=9cm]{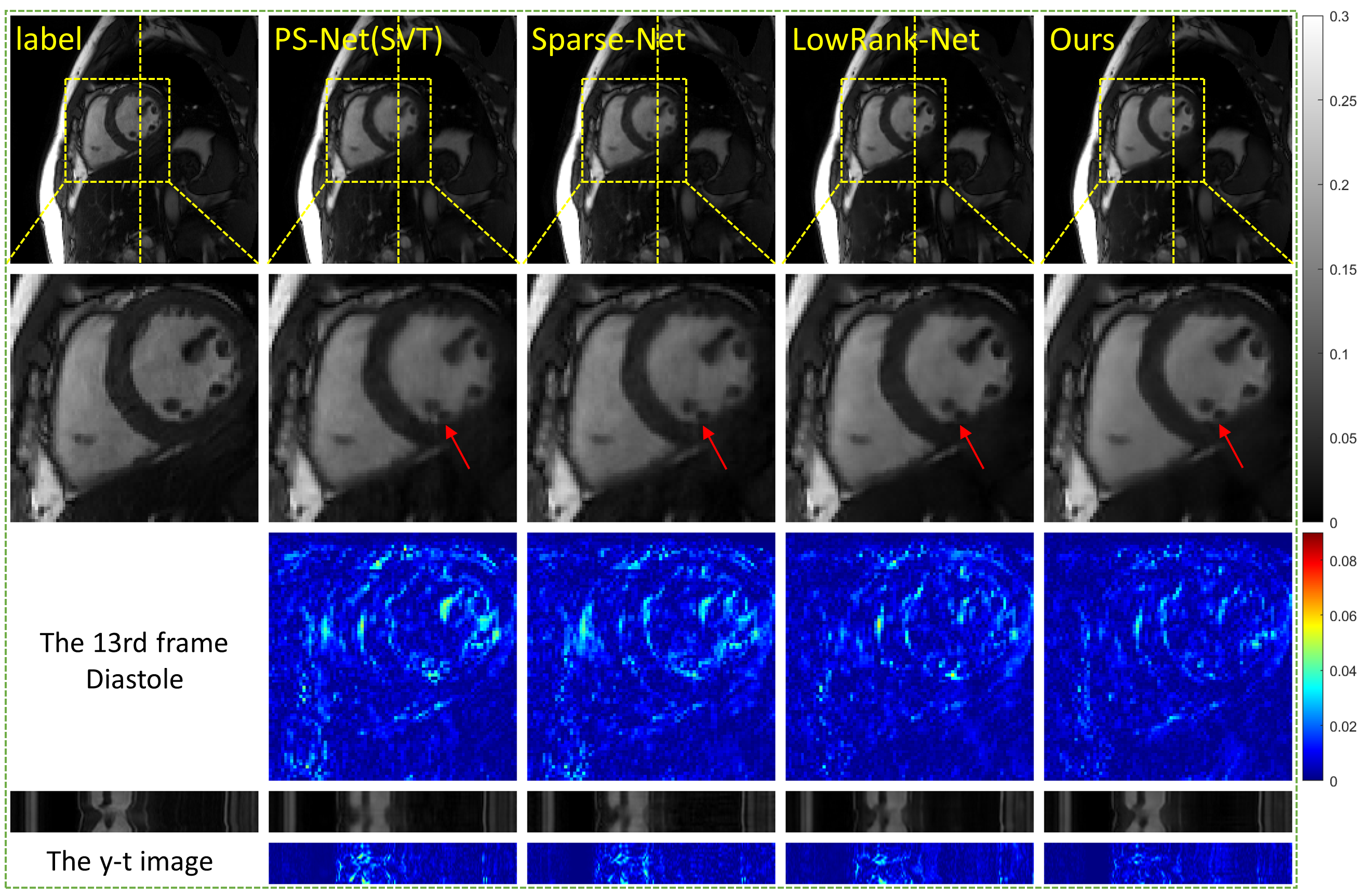}}
\caption{The ablation reconstruction results in multi-channel at 12-fold acceleration. The label and the reconstruction results for PS-Net(SVT), Sparse-Net, LowRank-Net, and PS-Net are displayed in the first row. The second row shows the zoomed-in view of the ROI region (indicated by the yellow box in the first row), and the third row shows the error map of the ROI region. The y-t image in the fourth row shows the 92nd slice along the $y$ and temporal direction. The fifth row shows the error map of the y-t image with an error range of [0--0.09].}
\label{Ablation}
\end{figure}

\begin{table}
\caption{The average quantitative evaluation metrics (MSE, PSNR, SSIM) of PS-Net(SVT), Sparse-Net, LowRank-Net and PS-Net in the multi-channel testing dataset when undersampled by 12-fold.}
\begin{center}
\begin{tabular}{c|cccc}
\hline \hline AF & Methods & MSE(*e-5) & PSNR (dB) & SSIM(*e-2) \\
\hline \multirow{4}{*}{12$\times$}&PS-Net(SVT) & 5.92 $\pm$ 3.13 & 42.81 $\pm$ 2.12 & 97.29 $\pm$ 1.07 \\
&Sparse-Net & 5.64 $\pm$ 2.86 & 42.98 $\pm$ 2.03 & 97.37 $\pm$ 0.98 \\
&LowRank-Net & 4.97 $\pm$ 2.36 & 43.47 $\pm$ 1.92 & 97.12 $\pm$ 1.02 \\
&PS-Net & \textbf{3.67 $\pm$ 1.89} & \textbf{44.87 $\pm$ 2.07} & \textbf{98.04 $\pm$ 0.85} \\
\hline \hline
\end{tabular}
\label{table SVT}
\end{center}
\end{table}

We designed ablation experiments to verify that adaptive low-rank learning strategy in PS-Net is effective. Three network were performed. The first one uses the SVT strategy to depict the low rank, implemented as the SVT module, which replaces the PS annihilation module, and the sparse module remains unchanged. This network is referred to as PS-Net (SVT). The second one removes the PS annihilation module, and the other modules and initialized hyper-parameters were kept unchanged, called Sparse-Net. The third one removes the sparse module and still uses the PS annihilation module for low-rank adaptive learning, named LowRank-Net. The initialization hyper-parameters of the above three networks are the same as PS-Net. The ablation experiments were performed in the multi-channel scenario with 12-fold acceleration. The reconstruction results are shown in Fig. \ref{Ablation}. PS-Net reconstructs the papillary muscle more clearly and more detailed than PS-Net (SVT), without artifacts. And the error map of PS-Net in third row and fifth row are clearer than PS-Net (SVT), which indicates that our adaptive low-rank learning strategy effectively portrays the low-rank of dynamic MR images more accurately. The reconstruction effect of LowRank-Net is better than that of PS-Net(SVT), Sparse-Net, which also shows that our adaptive low-rank learning strategy is effective. After adding the sparse prior, the reconstruction result of PS-Net achieves the best result among the four networks. Table \ref{table SVT} shows the quantitative metrics of the reconstruction.

\section{Discussion}

\subsection{Robustness experiments}

\begin{figure}
\centerline{\includegraphics[width=9cm]{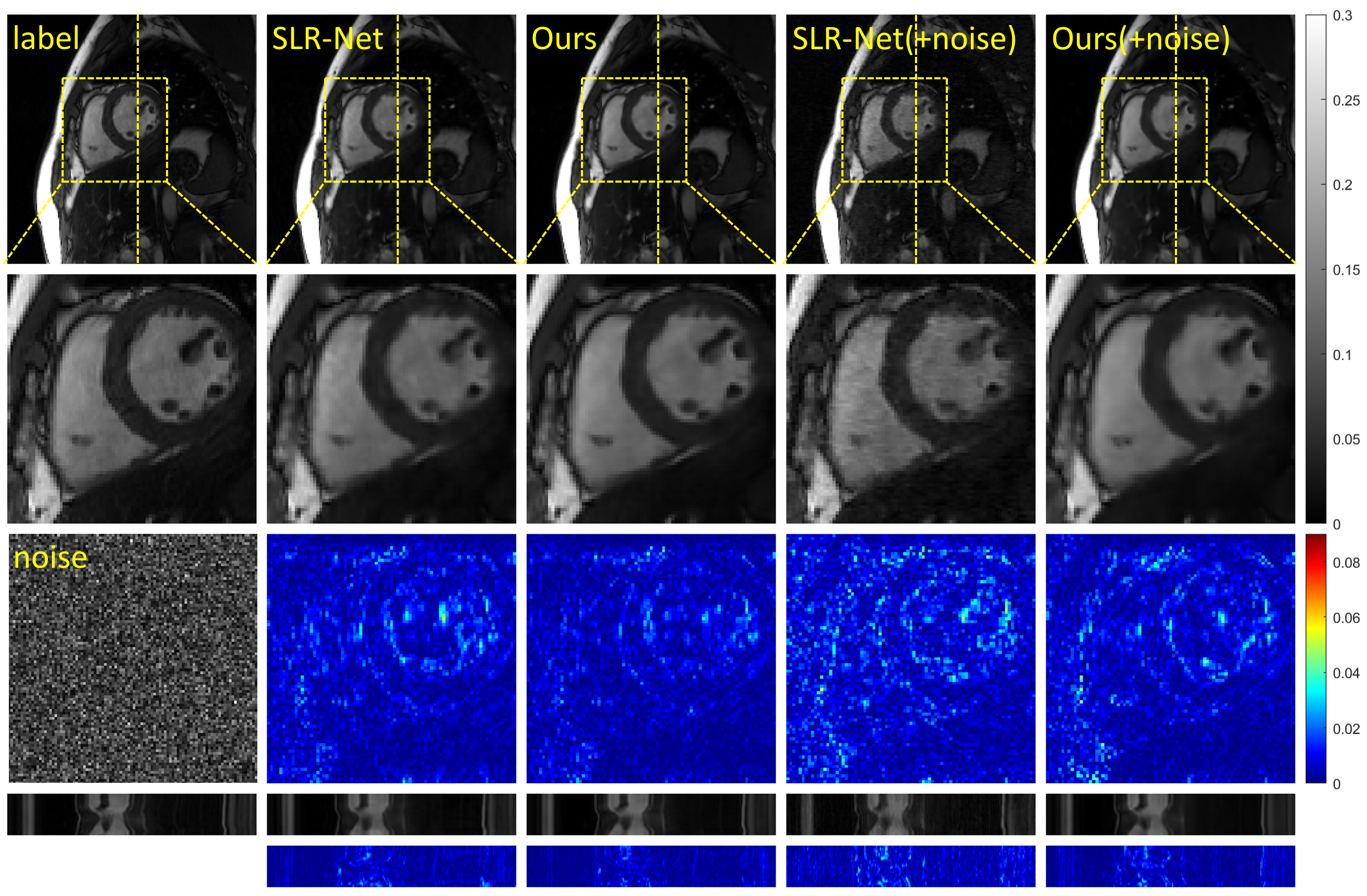}}
\caption{The reconstruction results after adding Gaussian white noise with variance 0.01 in the single-channel 8-fold acceleration scene. The intentionally added Gaussian white noise is in the third row. The label is in the first column. The reconstruction results of PS-Net compared to SLR-Net after adding or not Gaussian white noise are in the final four columns. In the first row, the labels and reconstruction results are displayed. The second row shows the zoomed-in view of the ROI region (indicated by the yellow box in the first row), and the third row shows the error map of the ROI region. The y-t image in the fourth row shows the 92nd slice along the $y$ and temporal direction. The fifth row shows the error map of the y-t image with an error range of [0--0.09]. Note that here the pixel amplitude ranges of the noise and the image are different.}
\label{noise results}
\end{figure}

\begin{figure*}
\centerline{\includegraphics[width=18.5cm]{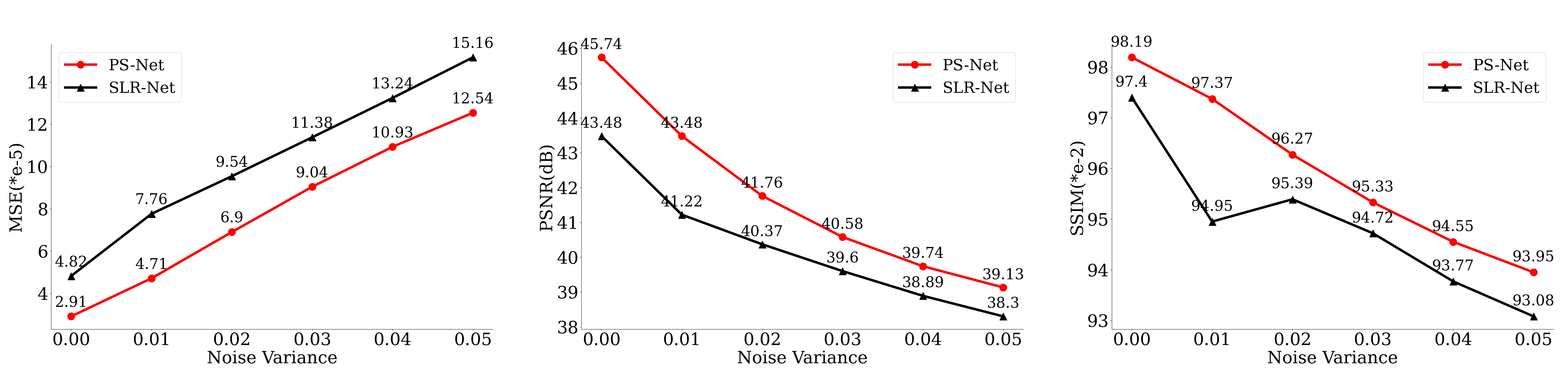}}
\caption{The robustness test results of PS-Net and SLR-Net. From left to right, show the curves of MSE, PSNR and SSIM of PS-Net and SLR-Net with the variation of Gaussian white noise intensity. The red curve is the result of PS-Net and the black curve is the result of SLR-Net.}
\label{curve}
\end{figure*}

\begin{table}
\caption{
The average quantitative evaluation metrics (MSE, PSNR, SSIM) of PS-Net in the multi-channel testing dataset when undersampled by 16-fold, 20-fold and 24-fold respectively.}
\begin{center}
\begin{tabular}{c|cccc}
\hline \hline Methods & AF & MSE(*e-5) & PSNR (dB) & SSIM(*e-2) \\
\hline \multirow{3}{*}{PS-Net}&16$\times$ & 6.72 $\pm$ 3.27 & 42.21 $\pm$ 2.04 & 97.08 $\pm$ 1.12 \\
&20$\times$ & 7.89 $\pm$ 4.06 & 41.54 $\pm$ 2.09 & 96.75 $\pm$ 1.27\\
&24$\times$ &9.36 $\pm$ 4.58 & 40.78 $\pm$ 2.01 & 96.25 $\pm$ 1.40 \\
\hline \hline
\end{tabular}
\label{table multicoil higher fold}
\end{center}
\end{table}

PS-Net represents the image by subspace, which can be seen as the generalization of PCA\cite{vaswani2018robust}. Hence PS-Net is more robust. The robustness of PS-Net is tested by artificially adding Gaussian white noise to the data during training and testing data. Fig. \ref{noise results} shows the reconstruction results after adding Gaussian white noise with a variance of 0.01\cite{variance}. The enlarged view of the second row in Fig. \ref{noise results} indicates that PS-Net sketches the papillary muscle boundary clearly and generates no artifacts in the core regions of the left and right ventricles, which are closer to the fully sampled label than SLR-Net. Besides, we conducted experiments with variance ranging from 0 to 0.05, and the quantitative metrics are shown in Fig. \ref{curve}. The quantitative metrics show that PS-Net suppress noise well even in the harsh case of noise variance of 0.05.

\subsection{Higher Acceleration Reconstruction}
\begin{figure}
\centerline{\includegraphics[width=9cm]{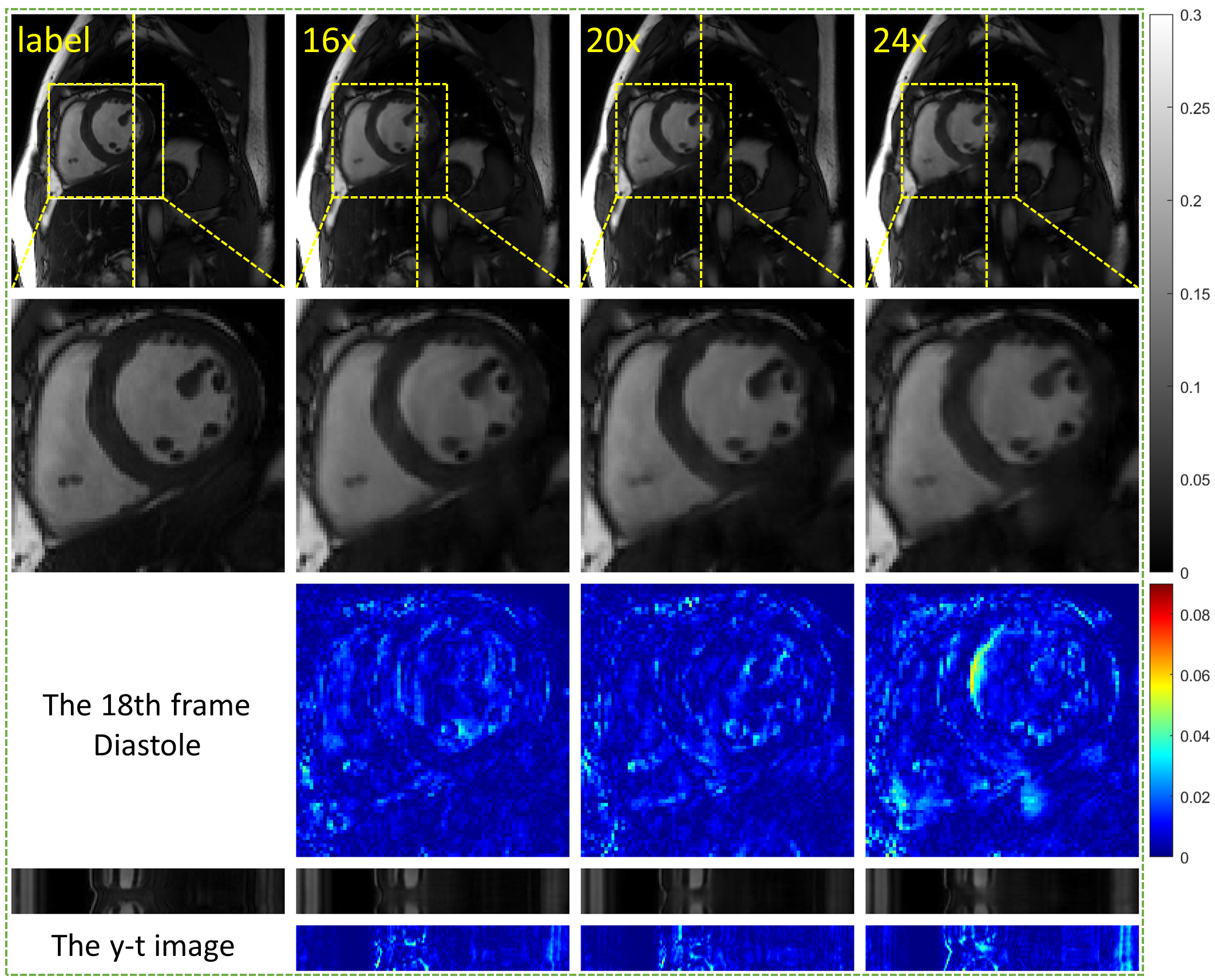}}
\caption{The reconstruction results of PS-Net in multi-channel scene at 16-fold, 20-fold and 24-fold acceleration. The label and the reconstruction results are displayed in the first row. 
The second row shows the zoomed-in view of the ROI region (indicated by the yellow box in the first row), and the third row shows the error map of the ROI region. The y-t image in the fourth row shows the 106th slice along the $y$ and temporal direction. The fifth row shows the error map of the y-t image row with an error range of [0--0.09].}
\label{multicoil higher fold}
\end{figure}

We also tested the performance of PS-Net at higher accelerations (16-fold, 20-fold, and 24-fold) with multi-coli data. The reconstruction results are shown in Fig. \ref{multicoil higher fold}, and the quantitative metrics are illustrated in Table \ref{table multicoil higher fold}. The enlarged view of the ROI region in the second row of Fig. \ref{multicoil higher fold} indicates that PS-Net can reconstruct the papillary muscle outlines even at 24-fold, and no artifacts are generated in either the left or right ventricular regions.

\subsection{Limitation}
Since our method exploits the equivalence of Hankel matrix multiplication and convolution, there are more convolutional layers in PS-Net than other model-driven reconstruction-based methods, resulting in the network needing to be trained with 10 million parameters. Besides, although PS-Net shortens the reconstruction time of dynamic MRI, its reconstruction time remains close to SLR-Net. It does not reflect the time advantage of avoiding SVD calculation. The reason is that PS-Net has more convolutional layers, which causes the computation to take more time, offsetting the time advantage of avoiding SVD computation.


\section{Conclusion}
This paper proposed a partially separable network (PS-Net) to more adaptively characterize the low-rank prior through a learnable null space transform. The null space transform was formulated as a Hankel matrix multiplication, equivalent to the convolution operation, and can be naturally unrolled into a convolutional network module. The ablation experiments show that the learnable low-rank representation enhances the low-rank characterization of the image than the nuclear norm and the reconstruction result is further improved by adding the sparse modules. Moreover, PS-Net has strong robustness and can suppress artifacts well. The experiments show that our method achieves the best reconstruction quality and quantitative evaluation metrics among the traditional compressed sensing method L+S, the deep learning end-to-end method CRNN, the model-driven based method MoDL and SLR-Net.

\section*{Acknowledgement}
This study was supported in part by the National Key R\&D Program of China nos. 2020YFA0712200, 2017YFC0108802, National Natural Science Foundation of China under grant nos. 62125111, 81830056,
U1805261, 61671441, 81971611, 81901736, and the Innovation and Technology Commission of the government of Hong Kong SAR under grant no. MRP/001/18X, the Key Laboratory for Magnetic Resonance and Multimodality Imaging of Guangdong Province under grant no. 2020B1212060051, the Guangdong Basic and Applied Basic Research Foundation no. 2021A1515110540, and by the Chinese Academy of Sciences program under grant no. 2020GZL006.

\bibliographystyle{IEEEtran}
\bibliography{refs}






\vfill

\end{document}